\documentclass{pasj01}
\Received{$\langle$reception date$\rangle$}
\Accepted{$\langle$acception date$\rangle$}
\Published{$\langle$publication date$\rangle$}

\usepackage{bm}
\usepackage{natbib}

\usepackage{color}

\newtheorem{algorithm}{Algorithm}

\newcommand{\V}{\bm}

\newlength{\figurewidth}
\figurewidth = \textwidth
\advance \figurewidth by -\columnsep
\divide  \figurewidth by 2

\newcommand{\eqref}[1]{(\ref{#1})}

\begin{document}

\title{PRECL: A new method for interferometry imaging from closure phase}
\author{Shiro \textsc{Ikeda}$^{1,2}$, 
        Fumie \textsc{Tazaki}$^3$, 
        Kazunori \textsc{Akiyama}$^{3,4,5}$,
        Kazuhiro \textsc{Hada}$^3$ $\&$
        Mareki \textsc{Honma}$^{3,6}$}
\altaffiltext{1}{The Institute of Statistical Mathematics, Tachikawa, Tokyo,
  190-8562, Japan}
\altaffiltext{2}{Department of Statistical Science, Graduate University for
  Advanced Studies (SOKENDAI), Tachikawa, Tokyo, 190-8562, Japan}
\altaffiltext{3}{Mizusawa VLBI Observatory, National Astronomical Observatory of Japan, Mitaka, Tokyo, 181-8588, Japan}
\altaffiltext{4}{Department of Astronomy, The
  University of Tokyo, Bunkyo-ku, Tokyo 113-0033, Japan}
\altaffiltext{5}{Massachusetts Institute of Technology, Haystack
  Observatory, Westford, MA 01886, USA}
\altaffiltext{6}{Department of Astronomical Science, Graduate University for
  Advanced Studies (SOKENDAI), Mitaka, Tokyo, 181-8588, Japan}
\email{shiro@ism.ac.jp}


\maketitle

\begin{abstract}
  For short-wavelength VLBI observations, it is difficult to measure
  the phase of the visibility function accurately. The closure phases are
  reliable measurements under this situation, though it is not
  sufficient to retrieve all of the phase information. We propose a
  new method, Phase Retrieval from Closure Phase (PRECL). PRECL
  estimates all the visibility phases only from the closure phases. Combining
  PRECL with a sparse modeling method we have already proposed,
  imaging process of VLBI does not rely on dirty image nor
  self-calibration. The proposed method is tested numerically and the
  results are promising. 
\end{abstract}

\section{Introduction}

Among observation methods of astronomy, Very Long Baseline
Interferometry (VLBI) achieves the highest angular resolution.
In particular, VLBI observations at short millimeter wavelength
($\lesssim 1.3$~mm) achieves unprecedentedly high spatial resolution
of few tens of microarcsecond, enabling to image shadows of
super-massive black holes in the center of our galaxy Sgr A* and nearby
radio galaxy M87. Although such observations have been technically
challenging due to the limited sensitivity of the instruments, fast
atmospheric phase fluctuations, and small number of available stations,
significant progresses have been made in the last several years with
the Event Horizon Telescope \citep[EHT;][]{Doeleman_etal.2008.nature,
  Doeleman_etal.2012.science,Fish_etal.2011.apjl,Lu_etal.2012.apjl,
  Lu_etal.2013.apj,Lu_etal.2014.apj,Akiyama_etal.2015.apj}.

In interferometry, observations at different stations are integrated
into a complex visibility function, which corresponds to the 
two-dimensional Fourier transform of the original image.
EHT has succeeded in measuring the visibility amplitude
\citep{Doeleman_etal.2008.nature,Doeleman_etal.2012.science,Fish_etal.2011.apjl}
and also the phase-related quantity through the closure phases
\citep{Lu_etal.2012.apjl,Lu_etal.2013.apj,Akiyama_etal.2015.apj,Fish_etal.2016.apj}.
But it is still difficult to obtain the visibility phases.

The image reconstruction from the visibility function is one of the
most important processes in interferometric observations. In particular,
high fidelity imaging in super-resolution regime is essential for EHT
observations of Sgr A* and M87, since the expected size of the black
hole shadow is comparable to or possibly smaller than the diffraction
limit of the telescope.
We have proposed a sparse modeling method for super resolution imaging
\cite[]{Honma_etal.2014.pasj}.
The method is robust against thermal noises and can reconstruct
high-fidelity images even in super-resolution regime. However, the
method might not be applicable to short-millimeter VLBI data, since
the complex information, in other words, phase information of the 
visibility function is required.
In this article, we propose a new method, Phase REtrieval from CLosure
phase (PRECL). PRECL recovers the phase information of visibility
function from closure phases \cite[]{Jennison.1958.mnras}.
Since closure phases are free from station-based phase errors that
are dominant in general, PRECL is robust against the
measurement errors of the visibility phases due to the fast atmospheric
fluctuation or delays in some instruments.
Combining PRECL with sparse modeling, an image is reconstructed
without any dirty image nor self-calibration
\cite[]{CornwellWilkinson.1981}. This idea is similar to BiSpectrum
Maximum Entropy Method \cite[BSMEM;][]{Buscher.1994.direct}, but PRECL
converges quickly by solving a 
pair of simple optimization problems iteratively.
Since the sparse modeling method for imaging is formulated as
a convex optimization problem, the whole imaging process of the
interferometry will be formulated as a combination of 
tractable simple optimization problems. We tested the proposed
method with two types of simulated data and the results are promising.

The rest of the paper is organized as follows. In Section
\ref{sec:PRECL}, the problem is mathematically formulated and the
details of PRECL are shown. We show the results of numerical
experiments in Section \ref{sec:experiments} and conclude the paper
after some discussions in Section \ref{sec:conclusion}.

\section{Phase retrieval from closure phase}
\label{sec:PRECL}

In this section, we show the mathematical formulation of the problem
and propose PRECL.

\subsection{Closure phase}

The goal of radio interferometric imaging is to reconstruct the
brightness distribution $I(x,y)$ of a target radio source at a
wavelength $\lambda$, where $(x,y)$ is a sky coordinate (in
the equatorial coordinate system) relative to the reference position
so-called the ``phase-tracking center.'' The observed quantity
$V(u,v)$ is a complex function called visibility, which is related to
$I(x,y)$ by the two-dimensional Fourier transform defined as
\begin{equation}
  V(u,v) = \int I(x,y) e^{-2\pi i(ux+vy)} dxdy.
\end{equation}
Here, the spatial frequency $(u,v)$ corresponds to the normalized
(with $\lambda$) baseline vector between two antennas projected to the
tangent plane of the celestial sphere at the phase-tracking
center. Since $I(x,y)$ is real, $V(u,v)$ has the Hermitian symmetry,
that is, $V(-u,-v) = V^{\ast}(u,v)$ where $^{\ast}$ denotes the
complex conjugate. The visibility is computed as 
the cross correlation of the input signals at two
stations. Thus, the number of the observed points in
$(u,v)$-plane is limited by the number of the antennas operating at
the same time.

Let $j$ be the index of measurements where the total number is
$N$: Corresponding position in $(u,v)$-plane is $(u_j, v_j)$,
recording time is $t_j$, and complex visibility
function is $V_j$. Let us define $V_j$ with its phase $\phi_j$ as
follows
\begin{equation}
 V(u_j,v_j) = V_j = |V_j|e^{i\phi_j},
\end{equation}
where, each phase satisfies $\phi_j\in(-\pi,\pi]$. 
In practical situations, instrumental delays and the atmospheric
turbulence mostly from the troposphere induce the antenna-based error
in the visibility phase. As the result, the observed phase 
$\tilde{\phi_j}$ is offset from the true phase $\phi_j$. 
This is a serious problem especially in VLBI 
observations performed at different sites with non-synchronized 
local oscillators.

The closure phase is defined as a combination of three
different visibility phases recorded at the same time and is known to be free
from the antenna-based phase errors \citep[]{Jennison.1958.mnras}. Let
us show the definition of the closure phase,
\begin{eqnarray*}
  \psi_m = \tilde{\phi}_j + \tilde{\phi}_k - \tilde{\phi}_l = \phi_j +
  \phi_k - \phi_l,
\end{eqnarray*}
where $m$ is the index of the closure phase. The closure phase has
been used to calibrate the visibility phases in the VLBI 
observations \citep{Rogers_etal.1974.apj}.

Let $\V{\phi}=(\phi_1,\cdots,\phi_N)^T$ and
$\V{\psi}=(\psi_1,\cdots,\psi_M)^T$ be the visibility and the closure 
phase vector, respectively. The above relation
is summarized as a system of linear equations,
\begin{equation}
  \label{eq:closure}
  \V{\psi} = A \V{\phi},
\end{equation}
where $A$ is a real $M\times N$ ($M<N$) matrix. The matrix is sparse
that each row has only three nonzero ($-1$ or $1$) components.  The
closure phase $\V{\psi}$ is computed from a set of
measurements.

\subsection{Outline of PRECL}

PRECL is a method to estimate $\V{\phi}$ from $\V{\psi}$. 
Since the system of linear equations in eq.~\eqref{eq:closure} is 
underdetermined, some additional information must be used.

We rely on an assumption that the phase $\phi_j$ does not behave
randomly but changes smoothly in $(u,v)$-plane. In other words, when
$(u_j,v_j)$ and $(u_k,v_k)$ are close to each other, $\phi_j$ and
$\phi_k$ are expected to be similar. Now, we consider minimizing the
following cost function under the constraint in eq.~\eqref{eq:closure}
\begin{equation}
  \label{eq:cost1}
  C(\V{\phi},\V{\xi}) = \frac12\sum_{j\ne k} w_{jk}
  (\phi_j-\phi_k-\xi_{jk})^2, 
\end{equation}
where the weight $w_{jk}$ is positive and symmetric
($w_{jk} = w_{kj}$). Each phase $\phi_j$ is in the interval
$(-\pi,\pi]$ and the variable $\xi_{ij}\in\{-2\pi,0,2\pi\}$ is
introduced in order to absorb the periodicity of the distance measure
between two phases. Now, PRECL is defined as an optimization problem.
\begin{eqnarray}
  &&\label{eq:PRECL_optimization}
  \min_{\V{\phi},\V{\xi}}C(\V{\phi},\V{\xi})\hspace{1em}
  \hspace{2em}\mbox{subject to}
  \hspace{2em}\V{\psi} = A \V{\phi},\\\nonumber
  &&\hspace{.1em}
  \V{\phi}\in(-\pi,\pi]^N,\hspace{.5em}
  \V{\xi}\in\{-2\pi,0,2\pi\}^{N\times (N-1)}.
\end{eqnarray}
This is the outline of PRECL. 

\subsection{Defining weights}

If one particular $w_{jk}$ is larger than other weights, corresponding $\phi_j$ and
$\phi_k$ should be close to make the cost function small. Thus, the
result of PRECL largely depends on the weights $\{w_{jk}\}$. Here
we discuss how to define them.

Relying on the assumption of smoothness, we define $w_{ij}$ with the
polar coordinate system. Let us define two variables as follows,
\begin{eqnarray}
  r_j = \sqrt{u_j^2+v_j^2},\hspace{1em}
  \displaystyle
  |\theta_{jk}| &=& \arccos\frac{|u_ju_k+v_jv_k|}{r_ir_k}. 
\end{eqnarray}
The weight $w_{jk}$ is defined with these variables as follows,
\begin{eqnarray}
\label{eq:def_w}
  w_{jk} = \alpha(r_j,r_k)\beta\bigl(|\theta_{jk}|\bigr).
\end{eqnarray}
We have tested different types of functions, and the following
combination of $\alpha$ and $\beta$ works best.
\begin{eqnarray}
\alpha(r_j,r_k) &=& \exp\Bigl(-\lambda_r \sqrt{|r_j^2-r_k^2|}\Bigr)\\
\beta\bigl(|\theta_{jk}|\bigr) &=& \exp\bigl(-\lambda_\theta
\sqrt{|\theta_{jk}|}\bigr)
\end{eqnarray}
where $\lambda_r$ and $\lambda_\theta$ are two tunable parameters. We
also restrict the number of non-zero components. 
We first initialize $w_{jk}=0$ for all $(j,k)$, then choose closest $D$ points from 
each $(u_j,v_j)$ and set those $w_{jk}$ positive according to
eq.~\eqref{eq:def_w}. This process is repeated for all $j$ and finally $w_{jk}$
is made symmetric.

\subsection{Solving the optimization problem}

Since $\V{\phi}$ is real and $\V{\xi}$ is discrete,
minimization of $C(\V{\phi},\V{\xi})$ is a mixed integer programming
(MIP) problem. MIP problems are difficult in general. 
Here, we propose a simple algorithm which converges quickly.
\begin{algorithm}
  Optimization algorithm for PRECL
  \begin{enumerate}
  \item Initialize $\V{\xi}$ and $\V{\phi}$ to a zero vector.
  \item Iterate (i) and (ii) alternately until convergence.
    \begin{enumerate}
      \renewcommand{\labelenumii}{(\roman{enumii}).}
    \item Fix $\V{\xi}$ and update $\V{\phi}$ by minimizing
      $C(\V{\phi},\V{\xi})$ over $\V{\phi}$.
    \item Fix $\V{\phi}$ and update $\V{\xi}$ by minimizing
      $C(\V{\phi},\V{\xi})$ over $\V{\xi}$.
    \end{enumerate}
  \end{enumerate}
\end{algorithm}

Each step of the iteration makes $C(\V{\phi},\V{\xi})$ smaller and
PRECL always converges. We explain steps {(i)} and {(ii)} in detail.
In step {(i)}, $\V{\xi}$ is fixed and $C(\V{\phi},\V{\xi})$ is
minimized.  Optimizing $C(\V{\phi},\V{\xi})$ over $\V{\phi}$ is
formulated as a quadratic programming (QP) problem. In order to see
this, let us define a matrix $H$ and a real vector $\V{f}$.
\begin{eqnarray*}
H &=& \{h_{jk}\}, \hspace{.5em}h_{jk} =
\left\{
\begin{array}{cc}
-w_{jk} & j\ne k\\
\sum_{l} w_{jl} & j = k
\end{array}
\right.,\\
\V{f} &=& (f_1,\cdots,f_N)^T,\hspace{1em}f_k=\sum_j w_{jk}\xi_{jk},
\end{eqnarray*}
where $^T$ denotes transpose.
$H$ is a real symmetric matrix and positive semi-definite. The
corresponding QP problem for $\V{\phi}$ is defined as follows
\begin{eqnarray}
  \label{eq:optimization}
  \min \Bigl[\frac12 \V{\phi}^T H \V{\phi} +\V{f}^T\V{\phi}\Bigr],
  \hspace{.2em}
  \mbox{subject to} 
  \hspace{.2em}
  A\V{\phi} = \V{\psi}, 
  \V{\phi}\in[-\pi,\pi]^N.
\end{eqnarray}
The above constraint of $\V{\phi}$ is different from
that in eq.~\eqref{eq:PRECL_optimization}. We solve the above QP
problem and treat the boundary condition separately if $\phi_i = -\pi$
for some $i$. The cost functions in $C(\V{\phi},\V{\xi})$ differs
from that in eq.~\eqref{eq:optimization} by the term which does not
include $\V{\phi}$. Thus, solving eq.~\eqref{eq:optimization} is equivalent to
minimizing $C(\V{\phi},\V{\xi})$ over $\V{\phi}$. There are many
solvers for QP problems\footnote{We used ``IBM ILOG CPLEX
  Optimizer.''}.

In step (ii), $\V{\phi}$ is fixed and $C(\V{\phi},\V{\xi})$ is
minimized over $\V{\xi}$. This is easily solved as follows,
\begin{eqnarray*}
\xi_{jk} =
\left\{
  \begin{array}{cl}
    2\pi, & \mbox{if }\phi_j-\phi_k > \pi\\
    0, & \mbox{if }-\pi< \phi_j-\phi_k \le \pi\\
    -2\pi, & \mbox{if }\phi_j-\phi_k \le -\pi\\
  \end{array}
\right..
\end{eqnarray*}
This is performed quickly.
Note that each step of PRECL converges quickly to its optimum and the
cost function decreases monotonically through iterations. It may
converges to a local minimum, especially if the phase structure is
very complicated.

\subsection{Sparse modeling method for image reconstruction}
\label{sec:sparsemodeling}

The complex visibility function $V(u,v)$ is reconstructed with the 
phases retrieved by PRECL. Next step is to reconstruct the intensity 
map $I(x,y)$. We have already proposed a super-resolution method for imaging in
\citet{Honma_etal.2014.pasj}. We explain the outline of the method and
an extension.

Let the reconstructed visibility vector
$\tilde{\V{V}}=(\tilde{V}_1,\cdots,\tilde{V}_{2N})^T$ where we treat
real and imaginary part separately as real components. Let the intensity map vector
$\V{I} = (I_1,\cdots,I_{L^2})^T$, where the size of the
image is $L\times L$. Ideally, $\tilde{\V{V}}$ and $\V{I}$ 
have the following linear relation
\begin{equation}
  \label{eq:Fourier}
  \tilde{\V{V}}=F\V{I},
\end{equation}
where $F\in \Re^{2N\times(L^2)}$ is a real Fourier matrix. Since
$2N<(L^2)$ holds in general for
VLBI, 
image reconstruction is an ill-posed problem. Also, the observations
include errors. We have proposed to use LASSO
\citep{Tibshirani.1996.jrssb} to solve the problem. In LASSO, the
following cost function is minimized,
\begin{eqnarray}
  \label{eq:lasso}
  \min \Bigl[\frac{1}{2}\|\tilde{\V{V}}-F\V{I}\|_{\ell_2}^2
    +\lambda_1\|\V{I}\|_{\ell_1}\Bigr],
  \hspace{.2em}
  \mbox{subject to} 
  \hspace{.2em}
  \V{I}\succeq 0.
\end{eqnarray}
Here $\|\cdot\|_{\ell_2}$ and $\|\cdot\|_{\ell_1}$ are $L2$ and $L1$
norm, respectively. Many algorithms have been proposed to solve
the problem. We used an iterative algorithm proposed in
\citet{BeckTeboulle.2009.siam}.

The above optimization problem works well for sparse celestial
objects. For less sparse objects, one possible solution is to add another
regularization term. A typical choice is the total-variation term
which is defined as,
\begin{eqnarray*}
TV(\V{I}) &=&
              \sum_{i=1}^{L-1}\sum_{j=1}^{L-1}
              \sqrt{(I_{i,j}-I_{i+1,j})^2+(I_{i,j}-I_{i,j+1})^2}\\
          &&+\sum_{i=1}^{L-1}|I_{i,L}-I_{i+1,L}|+\sum_{j=1}^{L-1}|I_{L,j}-I_{L,j+1}|,
\end{eqnarray*}
where $I_{i,j}$ is the $(i,j)$ pixel of the image. Now the extended
problem is
\begin{eqnarray}
  \label{eq:lasso2}
  \min \Bigl[\frac{1}{2}\|\tilde{\V{V}}-F\V{I}\|_{\ell_2}^2
    &&+\lambda_1\|\V{I}\|_{\ell_1}
    +\lambda_{TV}TV(\V{I})\Bigr],\\\nonumber
  \mbox{subject to} &&
  \hspace{.2em}
  \V{I}\succeq 0.
\end{eqnarray}
We call this extension LASSO+TV method. For the algorithm to solve this
problem, we used a modified algorithm of \citet{BeckTeboulle.2009.ieeeip}.

\section{Numerical experiments}
\label{sec:experiments}

We applied PRECL to two types of simulated data sets on
the radio galaxy M87 that is one of the most ideal sources to study
the nature of relativistic jets (e.g. see
\citealt{Hada_etal.2011.nature,Hada_etal.2012.apj,Hada_etal.2013.apj,
  Hada_etal.2014.apj} for a review) and also to probe the
event-horizon-scale structure of the super-massive black hole (BH)
(e.g. \citealt{Doeleman_etal.2012.science,Akiyama_etal.2015.apj}).

\subsection{Simulated data at 7mm}
\label{subsec:7mm}

\begin{figure*}[htb]
  \begin{center}
    \includegraphics[width=.9\textwidth]{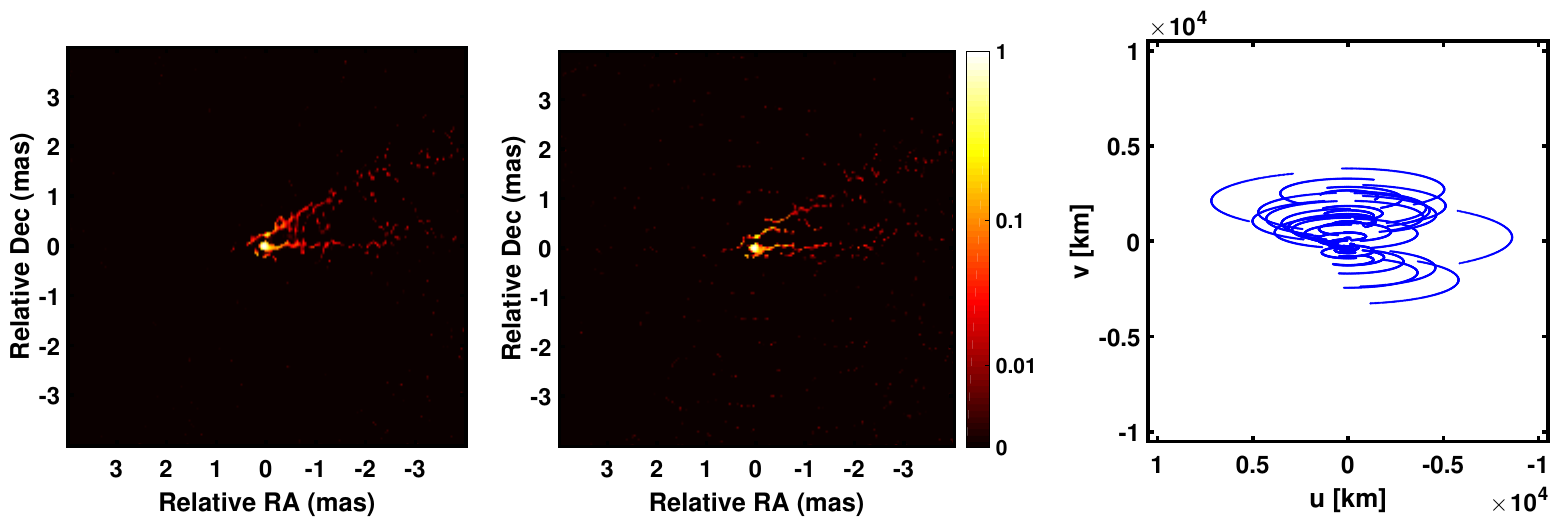}
  \end{center}
  \caption{ The data sets and the result of the first numerical
    experiment.
    The size of each images is 200$\times$200 pixels. Intensity is
    normalized with the maximum value.
    Left: Image of radio jet in M87 at 43 GHz reconstructed with full
    information of visibility.
    Center: Our result. Image of radio jet in M87 at 43 GHz reconstructed
    from closure phases by PRECL and LASSO.
    Right: $uv$-coverage of the measurement.  }
  \label{fig:Hada_clean}
\end{figure*}
The first data set is based on the results of Very Long Baseline Array
(VLBA) observations at 7~mm (43~GHz) presented in
\citet{Hada_etal.2011.nature}. Originally, observational data were
reduced in the Astronomical Image Processing System (AIPS) software
package of National Radio Astronomical Observatory (NRAO) and then
imaged in the DIFMAP software package \citep{Shepherd.1997} through
the iterative CLEAN and phase/amplitude self-calibrations (see
\citealt{Hada_etal.2011.nature} for details).

We simulated a VLBA observation at 7~mm with all the ten stations in
the AIPS task {\tt UVCON}. To make a simulated visibility data set of
M87, we imported the CLEAN model created in the above procedure into the
{\tt UVCON} task. We adopted a total bandwidth of 128~MHz and a
typical System Equivalent Flux Density (SEFD) of $\sim$1000~Jy to 
mimic a real VLBA observing condition
at 7~mm\footnote{see the status report of VLBA provided in the NRAO's
website https://science.nrao.edu/facilities/vlba}. The integration
time per simulated visibility was set to 150~seconds, assuming the
perfect phase coherence within the integration time. A continuous
observing run over 10~hours was assumed in this simulation.

In the millimeter VLBI observations, amplitude calibration can be
critical because systematic errors on measured baseline flux densities
reduce the imaging sensitivity in general. Here, we assumed the errors 
of flux density calibration were removed in the initial
calibration process. This reflects the realistic situation because,
at least at 7~mm observations with VLBA, errors can be mostly removed
with careful calibration
strategies. 

The simulated observation data were exported to the AIPS UV-FITS
format\footnote{%
see AIPS MEMO \#117
http://www.aips.nrao.edu/aipsmemo.html} 
with the AIPS task {\tt FITTP} and then imported into our software 
for PRECL. In the 
software, the closure phases were computed from the visibility 
phases and the original visibility phases were discarded. 
Then, the visibility phases were retrieved by PRECL from the visibility
amplitude in the UV-FITS file and the computed closure phases. We
used LASSO to create the image \citep{Honma_etal.2014.pasj} from the
original simulated data and the reconstructed data with PRECL.

In the left of Fig.~\ref{fig:Hada_clean}, the image of the M87 is
reconstructed with LASSO from the full visibility information. The image
in the center of Fig.~\ref{fig:Hada_clean} is reconstructed from the
closure phases and the amplitude information of the visibility function
using PRECL and LASSO. The $uv$-coverage of the
measurements are shown in the right. The
number of the points of the measurements was 11932 and that of the 
closure phases was 9258.

In order to apply PRECL to this data, we needed to define $\lambda_r$,
$\lambda_{\theta}$, and $D$. For this problem, we have the phase
calculated by self-calibration and we adjusted $\lambda_r$,
$\lambda_{\theta}$ and $D$ to make the results of PRECL close to those 
phases. As the results,
the parameters were set to $\lambda_r=6.67\times10^{-3}$,
$\lambda_{\theta}=12.0$ and $D=200$.
PRECL converges around 260~sec with a desktop computer (Intel core i7
CPU, Windows 10). The parameter $\lambda_1$ for LASSO is set to $1$
for both (left and center) of the Fig.~\ref{fig:Hada_clean}.

\begin{figure*}[htb]
  \begin{center}
    \includegraphics[width=.8\textwidth]{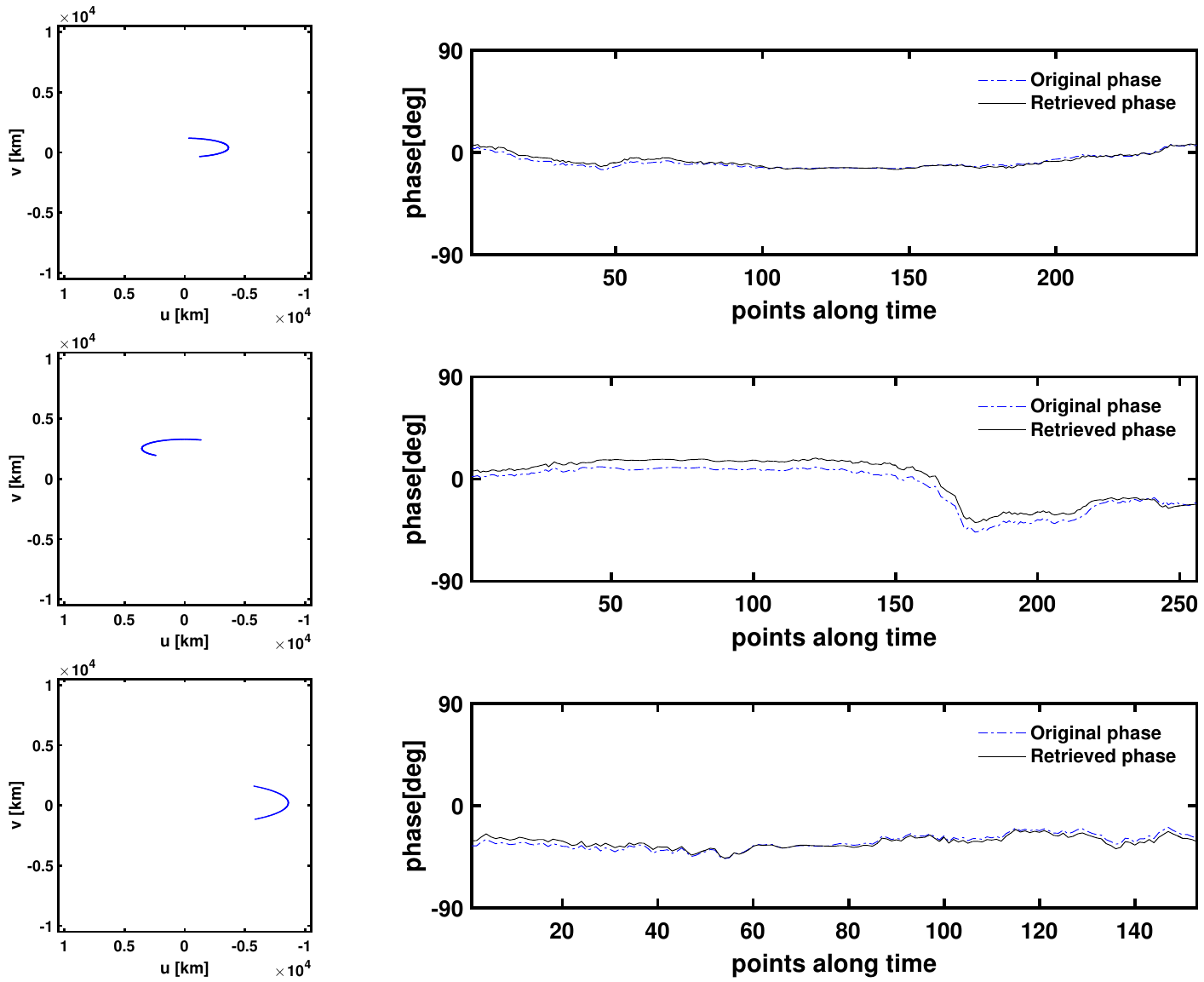}
  \end{center}
  \caption{The retrieved phases along the three $uv$-coverage curves. 
    Left: A curve corresponds to the $uv$-coverage of the
    station of VLBI. Right: Corresponding phases. Blue half dotted
    lines show the original phases created by iterative CLEAN and
    phase/amplitude self-calibration. Black solid lines show the
    phases retrieved by PRECL.}
   \label{fig:figure_PRECL_uv}
\end{figure*}
Let $\V{\phi}^\ast$ and $\hat{\V{\phi}}$ be the original and the
  estimated phase vector, respectively. In order to see the difference
  between $\V{\phi}^\ast$ and $\hat{\V{\phi}}$, we computed the
  following two functions.
\begin{eqnarray}
\label{eq:d1}
  d_1(\V{\phi}^\ast,\hat{\V{\phi}})
  &=\frac{1}{N}\sum_{i=1}^N\min
    \bigl(
    |\phi_i^\ast-\hat{\phi}_i|,2\pi-|\phi_i^\ast-\hat{\phi}_i|
    \bigr),\\
  \label{eq:d2}
  d_2(\V{\phi}^\ast,\hat{\V{\phi}})
  &=\frac{1}{N}\sum_{i=1}^N\min
    \bigl(
    |\phi_i^\ast-\hat{\phi}_i|,2\pi-|\phi_i^\ast-\hat{\phi}_i|
    \bigr)^2.
\end{eqnarray}
In the present example,
$d_1(\V{\phi}^\ast,\hat{\V{\phi}})=.0533$[rad] and
$d_2(\V{\phi}^\ast,\hat{\V{\phi}})=.00460$[rad$^2$]. %
Some of the retrieved phases are shown in
Fig.~\ref{fig:figure_PRECL_uv}. The phases are shown along three
$uv$-coverage curves. The results show that the phases are retrieved
almost perfectly. The performance does not have any frequency
dependency.

\subsection{Simulated data at 1.3 mm}
\label{subsec:1.3mm}

The second set of numerical experiments are based on physically
motivated models proposed for 1.3~mm emission from M87. We simulated
observations on the following four models based on different physical
assumptions.

The first model is a simple, but qualitatively correct force-free jet
model in magnetically dominated regime presented in
\citet{BroderickLoeb.2009.apj} and \citet{Lu_etal.2014.apj}. We
adopted a model image presented in \cite{Akiyama_etal.2015.apj}, which
is based on the model parameter fitted to the results of 1.3~mm
observations with EHT in \citet{Doeleman_etal.2012.science} and the
Spectral Energy Distribution (SED) of M87 (Broderick et al. in
preparation). The approaching jet is predominant for this model (see
\citealt{Akiyama_etal.2015.apj}) and we call it BL09 
(``Broderick \& Loeb 2009'' model).

The second and third ones are based on results of General Relativistic
Magnetohydrodynamic (GRMHD) simulations presented in
\citet{Dexter_etal.2012.mnras}. We use the representative models DJ1
and J2 in \citet{Dexter_etal.2012.mnras}, which are based on the same
GRMHD simulation but with different energy and spatial distributions
for radio-emitting leptons. The dominant emission region is the
accretion disk in DJ1 ({``disk + counter jet''} model) and the counter
jet in J2 ({``counter jet''} model) illuminating the last photon
orbit. We adopt model images in \cite{Akiyama_etal.2015.apj} where the
position angle of the large-scale jet for models is adjusted to
$-70^\circ$ inferred for M87 (e.g. \citealt{Hada_etal.2011.nature}).

The last model is based on results of GRMHD simulations presented in
\citet{Moscibrodzka_Falcke_Shiokawa.2016.aap}. We use the image
which is time-averaged for $\sim$ three months for our simulation. The image
has a dominant contribution from the counter jet illuminating the last
photon orbit similar to J2 of \citet{Dexter_etal.2012.mnras} but the
model assumes energy and spatial distributions of leptons very
different to J2. We rotated original images in
\cite{Moscibrodzka_Falcke_Shiokawa.2016.aap} to adjust the position
angle of the large-scale jet to $-70^\circ$. We call this model
MF15 
(``Mo{\'s}cibrodzka \& Falcke 2015'' model).

The simulated observation data with EHT at 1.3~mm (230~GHz) were
generated using the MIT Array Performance Simulator (MAPS)
package\footnote{http://www.haystack.mit.edu/ast/arrays/maps/} for
each of the above four models. The conditions of simulations followed
those in \cite{Lu_etal.2014.apj}. We adopted a band width of 8~GHz in
each of two polarizations which corresponds to a bit rate of
64~Gbit~s$^{-1}$ for four-level signals sampled at the Nyquist
rate. This was the target for the Atacama Large
Millimeter/submillimeter Array (ALMA) beam former
\citep{Fish_etal.2013.arxiv}. We adopted an integration time of 
120~s. The simulated observations were performed during 24~hours, and
the visibility was sampled at 6~minutes scans with an interval of
every 20~minutes when elevations of each station is higher than
fifteen degrees.

We assumed perfect phase coherence within the integration
time. Typical atmospheric coherence times at 230~GHz are 10~s, but it
varies from a few to few tens of seconds depending on the weather
condition at each observing site
\citep{Doeleman_etal.2009.apj}. Although the phase fluctuation due to
atmospheric turbulence does not affect the closure phase, such short
coherence time can cause serious coherent losses in the
coherently-averaged visibility amplitude. However, in practice, the
influence of the coherence losses can be removed with established
algorithms of the incoherent averaging
\citep{Rogers_Doeleman_Moran.1995.aj} and indeed have been calibrated
in the previous observations \citep{Doeleman_etal.2008.nature,
  Doeleman_etal.2012.science,Fish_etal.2011.apjl,Lu_etal.2012.apjl,
  Lu_etal.2013.apj,Akiyama_etal.2015.apj}.  We assumed the coherence
loss on the visibility amplitude was corrected as in the previous
observations.

For the other conditions of the simulation, we followed 
\cite{Lu_etal.2014.apj}. The array used for the simulation consists 
of stations at 8 different sites: 
Hawaii, consisting of one or more of the 
James Clark Maxwell Telescope (JCMT) 
and 
Submillimeter Array (SMA) 
phased together into a single
aperture;
the 
Arizona Radio Observatory Submillimeter Telescope (SMT) 
on Mount Graham;
the Combined Array for Research in Millimeter-wave Astronomy (CARMA) 
site in California; 
the Large Millimeter Telescope (LMT) on Sierra Negra, Mexico; 
the ALMA; 
the Institut de Radioastronomie Millim\'etrique (IRAM) 30~m
telescope on Pico Veleta (PV), Spain; 
the IRAM Plateau de Bure
Interferometer (PdBI), phased together as a single aperture; 
and 
the Greenland telescope (GLT). 
The flux density for each station was assumed to be equivalent as in
\cite{Lu_etal.2014.apj}.

Amplitude calibration is critical also at 1.3~mm. Here, we do not
include flux density errors as in the case of 7~mm. They were assumed
to be removed with existing self-calibration techniques using
different types of information, such as, measurements between
different bands, polarizations information, and the same
sites \citep{Fish_etal.2011.apjl,Lu_etal.2012.apjl,Lu_etal.2013.apj,
  Akiyama_etal.2015.apj}, and with new techniques using the suitable
amplitude calibrator or the closure amplitude information. The problem
of these errors will be considered carefully in the future work.

\begin{figure}[htbp]
  \begin{center}
    \includegraphics[width=0.6\figurewidth]{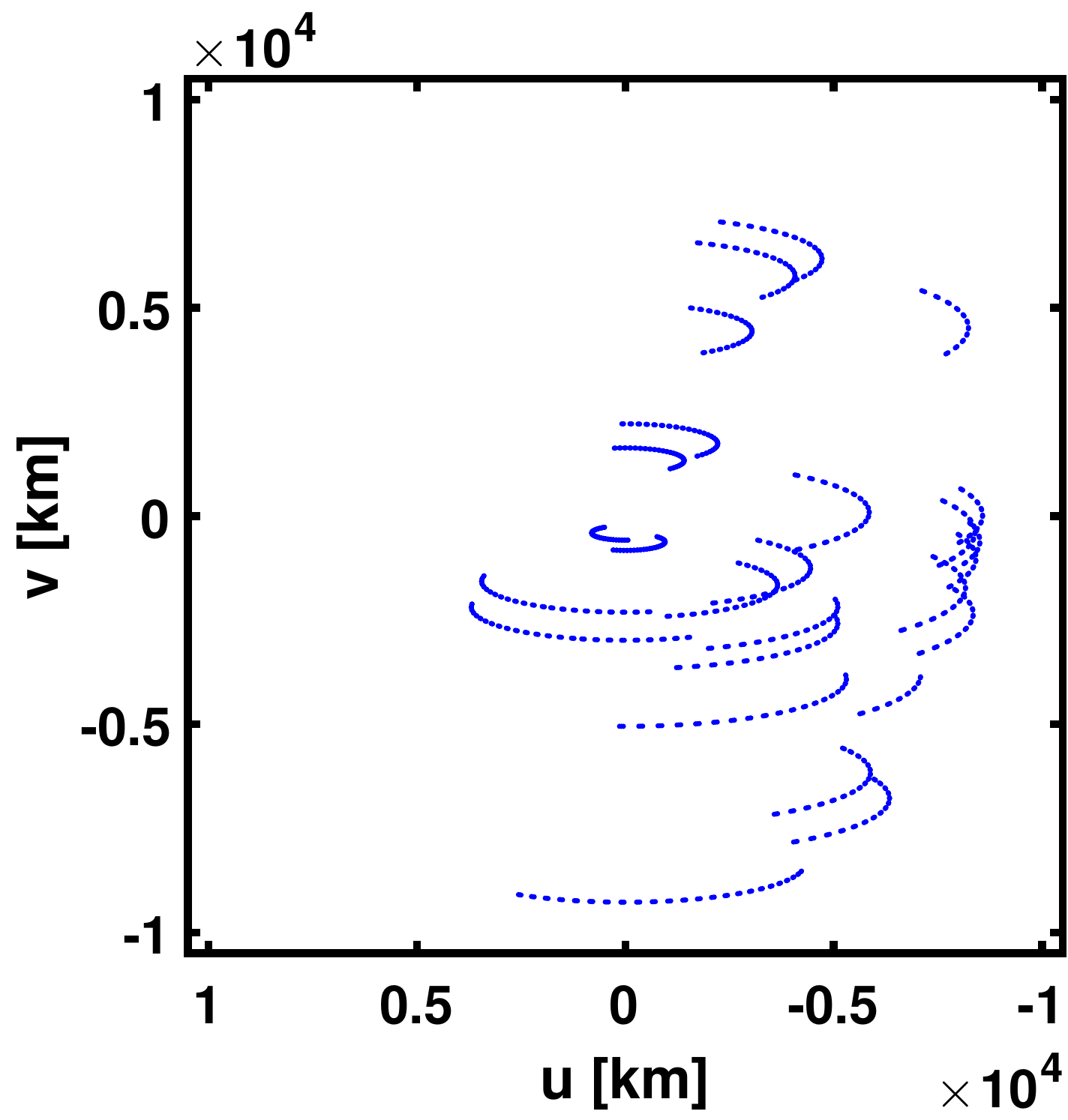}
  \end{center}
  \label{fig:simulation_original}
  \caption{Simulated EHT $uv$-coverage.}
\end{figure}

The number of the $uv$-coverage points was 1446 while the closure
phases were computed for 891 points
(Fig.~\ref{fig:simulation_original}). The simulated observational
data were converted to the AIPS UV-FITS format with the MAPS task {\tt
  maps2uvfits} and then imported into our software for PRECL similarly
to the previous 7~mm data. We used LASSO
($\lambda_1=1$) to reconstruct images from the original
simulated data, where visibility phases were given. 
For the reconstruction with the phases retrieved by
PRECL, we used LASSO+TV ($\lambda_1 = 15$
and $\lambda_{TV} = 2$). The parameters were set to
$\lambda_r=5.17\times 10^{-3}$, $\lambda_{\theta}=11.0$, and $D=70$. The
computational time for PRECL was low. It converged with less than 4 
iterations of steps (i) and (ii) where the computational time was within
6~sec (Intel core i7 CPU, Windows 10).

The results of PRECL are summarized in
Tab.~\ref{tab:results_precl}. For each model, the cost function of
PRECL is calculated with the original phase $\V{\phi}^\ast$ and the
estimated phase $\hat{\V{\phi}}$ and the difference between
$\V{\phi}^\ast$ and $\hat{\V{\phi}}$ is measured with
$d_1(\cdot,\cdot)$ and $d_2(\cdot,\cdot)$ defined in
eqs.~\eqref{eq:d1} and \eqref{eq:d2}, respectively.

\begin{table}[hbt]
  \centering
  \caption{
      The results of PRECL applied for simulated observations 
      at 1.3~mm: $\V{\phi}^\ast$ and $\hat{\V{\phi}}$ are the 
      original and the estimated phase, respectively. The cost functions for 
      $\V{\phi}^\ast$ and $\hat{\V{\phi}}$ are shown in the first and
      the second row.
      The difference between $\V{\phi}^{\ast}$ and $\hat{\V{\phi}}$ is 
      summarized in the third and the last row with $d_1(\cdot,\cdot)$ 
      and $d_2(\cdot,\cdot)$, respectively.
    }
  \label{tab:results_precl}
  \begin{tabular}{l||l|l|l|l}
    & BL09 & DJ1 & J2 & MF15 \\\hline\hline
    \rule[-1.8ex]{0pt}{5ex}
    $C(\V{\phi}^\ast,\V{\xi}^\ast)$
    & 1.920 & 0.890 & 0.801 & 0.718\\\hline
    \rule[-1.8ex]{0pt}{5ex}
    $C(\hat{\V{\phi}}, \hat{\V{\xi}})$
    & 0.574 & 0.326 & 0.262 & 0.425\\\hline
    \rule[-1.8ex]{0pt}{5ex}
    $d_1(\V{\phi}^\ast,\hat{\V{\phi}})$ [rad]   
    & 0.396 & 0.175 & 0.204 & 0.076\\\hline
    \rule[-1.8ex]{0pt}{5ex}
    $d_2(\V{\phi}^\ast,\hat{\V{\phi}})$ [rad$^2$]
    & 0.246 & 0.053 & 0.066 & 0.010\\\hline
  \end{tabular}
\end{table}

\begin{figure*}[hbt]
  \begin{center}
    \includegraphics[width=\textwidth]{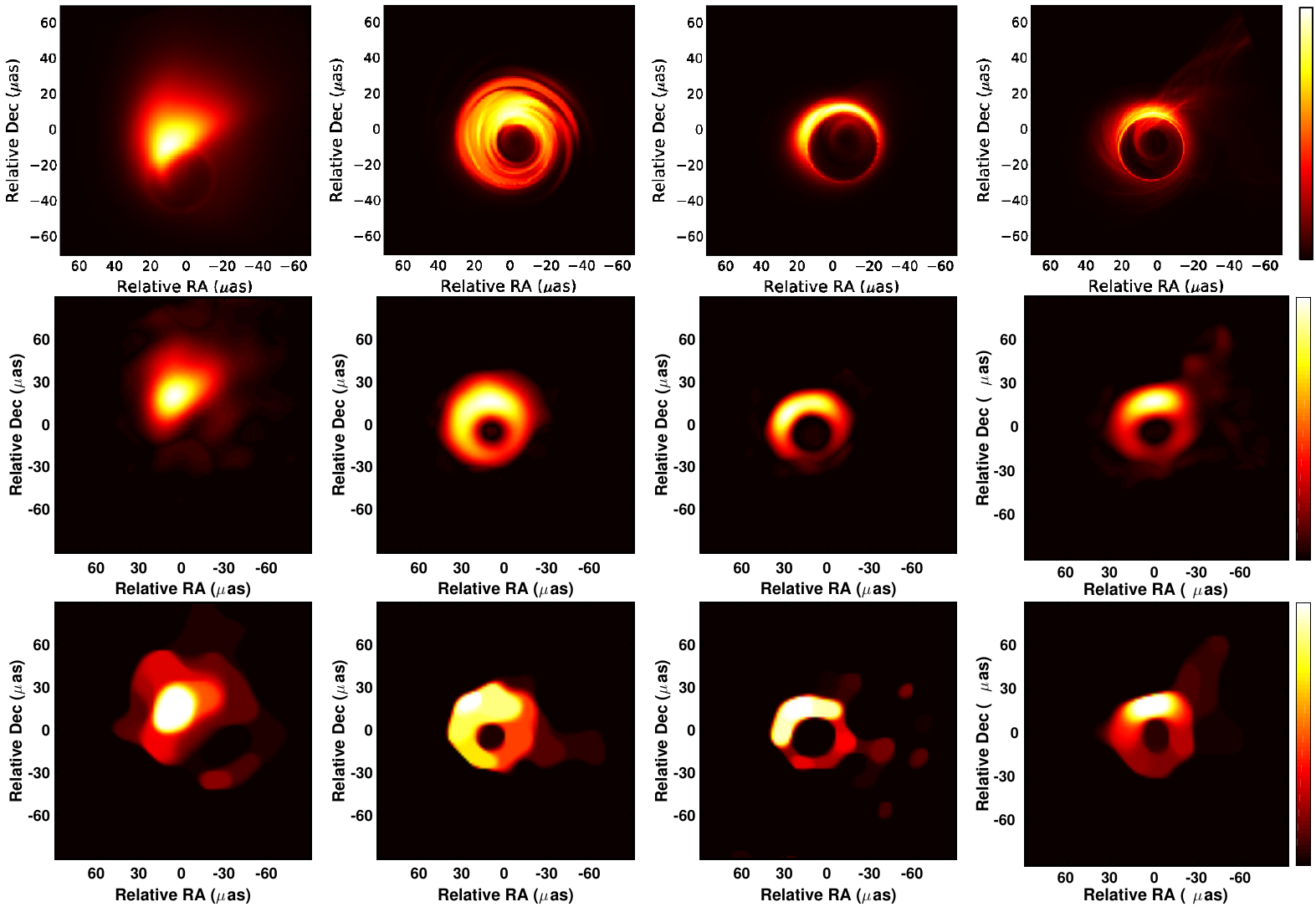}
  \end{center}
  \caption{The reconstructed images. Top row:
    Original images used for the simulation (\citet{Akiyama_etal.2015.apj} (left 3 images) and \citet{Moscibrodzka_Falcke_Shiokawa.2016.aap})
    Middle row: images
    reconstructed from full visibility information with LASSO. Image
    resolutions are 100$\times$100 pixels.
    Bottom row: images reconstructed
    from closure phases with PRECL and LASSO+TV. Image resolutions are
    100$\times$100 pixels. From left to right:
    1. {\em{BL09}} model in \citet{BroderickLoeb.2009.apj} 
    and \citet{Lu_etal.2014.apj}, 
    2. {\em{DJ1}} model in \citet{Dexter_etal.2012.mnras}, 
    %
    3. {\em{J2}} model in \citet{Dexter_etal.2012.mnras}, 
    %
    and 
    4. {\em{MF15}} model in \cite{Moscibrodzka_Falcke_Shiokawa.2016.aap}. 
    \label{fig:simulation}}
\end{figure*}

The reconstructed images for four
models are shown in Fig.~\ref{fig:simulation}. The reconstructed images
with phase information are compared with the reconstructions only with
closure phases. The reconstructions from closure phases are not
perfect, but the characteristics of the BH shadows are well preserved.

\section{Discussion}
\label{sec:conclusion}

  The proposed approach was applied for two types of data. 
  First data set was the simulated 7~mm observation (sec \ref{subsec:7mm}). 
  In this case, a small and bright celestial object is located in the 
  center, and the phase is rather flat. The retrieved phases are almost 
  perfect (Fig.~\ref{fig:figure_PRECL_uv}), and the quality of the 
  reconstructed image is very close to the original image 
  (Fig.~\ref{fig:Hada_clean}).

The other data sets were the simulated 1.3~mm observations 
  (sec \ref{subsec:1.3mm}). We tested four
  models. The phase structures are different depending on the model, but
  all of them have more complicated structures than 7~mm observation 
  because of the BH shadow.
  The results of PRECL are summarized in
  Tab.~\ref{tab:results_precl}. 
  The first and the second row of the table show the values of the
  cost function. These values reflect the difficulty of the
  problem. If the phase of the visibility function is not flat but has
  some structure, the cost function tends to be large. 
  The cost function with the estimated phase is smaller than that with
  the original phase. This is not surprising since the function was
  not optimized for the original phase.
  Although the differences between $\V{\phi}^\ast$ and
  $\hat{\V{\phi}}$ are not as small as the 7~mm simulation, they are
  good enough that the reconstructed images preserve the
  characteristics of the BH shadows very well
  (Fig.~\ref{fig:simulation}).

In the above numerical experiments, we treated observational/simulated
data with relatively small number of stations, i.e., $10$ for the VLBA 
case and $8$ for the EHT case. In an ideal case, the number of
independent closure phases for a snap shot observation is given by
\begin{equation}
N_{closure} = \frac{1}{2}(N_{ant}-1)(N_{ant}-2),
\end{equation}
where $N_{ant}$ is the number of the independent stations. The number
of baselines is
\begin{equation}
N_{base} = \frac{1}{2}N_{ant} (N_{ant}-1).
\end{equation}
Therefore, the ratio of the closure phases over the baseline phases,
which we call ``closure filling factor'' is given by
\begin{equation}
R_{closure} =
\frac{N_{closure}}{N_{base}}
=
\frac{N_{ant}-2}{N_{ant}}
.
\end{equation}
Here we assumed that the source is always visible from all the
stations. Practically however, some stations could be partially
dropped off due to elevation limit depending on the location of the
station as well as the source position, and the actual value of
$R_{closure}$ would be somewhat smaller than that for the ideal
case. For instance, the EHT simulation above has
$R_{closure} = 891/1446 \sim 0.62$, while the ideal case for
$N_{ant} = 8$ gives $R_{closure} = 0.75$.

If one considers an interferometric array with a larger number of
stations, the closure filling factor becomes higher and even close to
unity. For instance, in the case of JVLA ($N_{ant} = 27$) the closure
filling factor is $0.93$, and in the case of ALMA ($N_{ant} \sim 50$) the
factor can be as high as $0.96$. In these cases, $R_{closure}$ is
nearly unity, and hopefully visibility phases can be retrieved from
closure phases with even better accuracy than the results shown
above. We note that this would bring a significant impact in
calibration and imaging with interferometer: in the interferometric
imaging, the key calibration process is that concerning the phase,
which is directly affected by the fluctuation of troposphere (and
ionosphere as well in case of low frequency observations). Since the
closure phase totally cancels out station-based phase fluctuation
(both that of troposphere and ionosphere), an introduction of the
technique like PRECL can fairly simplify the interferometric
calibration process. The trade-off is that if one wants to apply this
technique to the interferometry with a large number of stations, it
requires large computational costs. This will be an issue for future
to be solved by high-speed parallel computing and/or algorithm
optimization.

\section{Summary}

We have proposed a new method for phase retrieval. The proposed PRECL
is different from the BSMEM
\cite[]{Buscher.1994.direct,Katagiri.1997.pasj}. The phase retrieval
problem is separated from the image reconstruction
and solved efficiently with a simple algorithm which always converges.
%
%
PRECL may converge to a local optimum, but we have not seen any
serious problem throughout the numerical experiments in this paper. We
will test PRECL with more complicated images to see its general
performance. This is one of our future problems.

We have combined PRECL with sparse modeling methods in order to
reconstruct the images. 
  We have tested this combination with simulated data sets. For the
  simulated BH observations with 7mm, the reconstruction is almost
  perfect. For 1.3~mm wavelength, the characteristics of the BH
  shadows are well preserved. 
%
  Although the quality of the reconstructed images may not be perfect,
  PRECL is simple, and converges quickly.  Therefore, the estimated
  phases are good enough for the initial values of the phases for
  iterative CLEAN, or MEM based method.

\section*{Acknowledgment}

We would like to thank Drs. A. E.~Broderick, J.~Dexter and
M.~Mo{\'s}cibrodzka for providing the data for numerical simulations.
We are grateful to the referee, Prof. S. K. Okumura for her
constructive suggestions. Finally, we thank the editor, Prof. Tsuboi
for his assistance with this submission. This study has been supported
by JSPS KAKENHI Grant Number 25120007 and 25120008. KA and KH are
supported by a Grant-in-Aid for Research Fellows of the Japan Society
for the Promotion of Science (JSPS).


\begin{thebibliography}{}
\expandafter\ifx\csname natexlab\endcsname\relax\def\natexlab#1{#1}\fi

\bibitem[{Akiyama {et~al.}(2015)Akiyama, Lu, , Fish, Doeleman, Avery
  E.~Broderick, Hada, Kino, Nagai, Honma, Johnson, Algaba, Asada, Brinkerink,
  Blundell, Bower, Cappallo, Crew, Dexter, Dzib, Freund, Friberg, Gurwell, Ho,
  Inoue, Krichbaum, Loinard, MacMahon, Marrone, Moran, Nakamura, Nagar,
  Ortiz-Leon, Plambeck, Pradel, Primiani, Rogers, Roy, SooHoo, Tavares,
  Tilanus, Titus, Wagner, Weintroub, Yamaguchi, Young, Zensus, \&
  Ziurys}]{Akiyama_etal.2015.apj}
Akiyama, K., Lu, R.-S., Fish, V.~L, {et~al.} 2015, \apj, 870, 150

\bibitem[{Beck \& Teboulle(2009a)}]{BeckTeboulle.2009.ieeeip}
Beck, A., \& Teboulle, M. 2009a, IEEE trans. Image Proc., 18, 2419

\bibitem[{Beck \& Teboulle(2009b)}]{BeckTeboulle.2009.siam}
Beck, A., \& Teboulle, M. 2009b, SIAM J. on Imaging Sci., 2, 183

\bibitem[{Broderick \& Loeb(2009)}]{BroderickLoeb.2009.apj}
Broderick, A.~E., \& Loeb, A. 2009, \apj, 697, 1164

\bibitem[{Buscher(1994)}]{Buscher.1994.direct}
Buscher, D. 1994, in Very High Angular Resolution Imaging (Springer
  Netherlands), 91--93

\bibitem[{Cornwell \& Wilkinson(1981)}]{CornwellWilkinson.1981}
Cornwell, R., \& Wilkinson, P.~N. 1981, \mnras, 196, 1067

\bibitem[{Dexter {et~al.}(2012)Dexter, McKinney, \&
  Agol}]{Dexter_etal.2012.mnras}
Dexter, J., McKinney, J.~C., \& Agol, E. 2012, \mnras, 421, 1517

\bibitem[{Doeleman {et~al.}(2009)Doeleman, Fish, Broderick, Loeb, \&
  Rogers}]{Doeleman_etal.2009.apj}
Doeleman, S.~S., Fish, V.~L., Broderick, A.~E., Loeb, A., \& Rogers, A.~E.
  2009, \apj, 695, 59

\bibitem[{Doeleman {et~al.}(2008)Doeleman, Weintroub, Rogers, Plambeck, Freund,
  Tilanus, Friberg, Ziurys, Moran, Corey, {et~al.}}]{Doeleman_etal.2008.nature}
Doeleman, S.~S., Weintroub, J., Rogers, A.~E., {et~al.} 2008, \nat, 455, 78

\bibitem[{Doeleman {et~al.}(2012)Doeleman, Fish, Schenck, Beaudoin, Blundell,
  Bower, Broderick, Chamberlin, Freund, Friberg,
  {et~al.}}]{Doeleman_etal.2012.science}
Doeleman, S.~S., Fish, V.~L., Schenck, D.~E., {et~al.} 2012, Science, 338, 355

\bibitem[{Fish {et~al.}(2013)Fish, Alef, Anderson, Asada, Baudry, Broderick,
  Carilli, Colomer, Conway, Dexter, {et~al.}}]{Fish_etal.2013.arxiv}
Fish, V., Alef, W., Anderson, J., {et~al.} 2013, arXiv:1309.3519

\bibitem[{Fish {et~al.}(2011)Fish, Doeleman, Beaudoin, Blundell, Bolin, Bower,
  Chamberlin, Freund, Friberg, Gurwell, {et~al.}}]{Fish_etal.2011.apjl}
Fish, V.~L., Doeleman, S.~S., Beaudoin, C., {et~al.} 2011, \apjl, 727, L36

\bibitem[Fish {et~al.}(2016)]{Fish_etal.2016.apj} Fish, V.~L.,
  {et~al.}, 2016, \apj, in press, arXiv:1602.05527

\bibitem[{Hada {et~al.}(2011)Hada, Doi, Kino, Nagai, Hagiwara, \&
  Kawaguchi}]{Hada_etal.2011.nature}
Hada, K., Doi, A., Kino, M., {et~al.} 2011, \nat, 477, 185

\bibitem[{Hada {et~al.}(2012)Hada, Kino, Nagai, Doi, Hagiwara, Honma,
  Giroletti, Giovannini, \& Kawaguchi}]{Hada_etal.2012.apj}
Hada, K., Kino, M., Nagai, H., {et~al.} 2012, \apj, 760, 52

\bibitem[{Hada {et~al.}(2013)Hada, Kino, Doi, Nagai, Honma, Hagiwara,
  Giroletti, Giovannini, \& Kawaguchi}]{Hada_etal.2013.apj}
Hada, K., Kino, M., Doi, A., {et~al.} 2013, \apj, 775, 70

\bibitem[{Hada {et~al.}(2014)Hada, Giroletti, Kino, Giovannini, D'Ammando,
  Cheung, Beilicke, Nagai, Doi, Akiyama, {et~al.}}]{Hada_etal.2014.apj}
Hada, K., Giroletti, M., Kino, M., {et~al.} 2014, \apj, 788, 165

\bibitem[{Honma {et~al.}(2014)Honma, Akiyama, Uemura, \&
  Ikeda}]{Honma_etal.2014.pasj}
Honma, M., Akiyama, K., Uemura, M., \& Ikeda, S. 2014, \pasj, 66, 95

\bibitem[{Jennison(1958)}]{Jennison.1958.mnras}
Jennison, R. 1958, \mnras, 118, 276

\bibitem[{Katagiri {et~al.}(1997)Katagiri, Morita, Kawaguchi, \&
  Hayakawa}]{Katagiri.1997.pasj}
Katagiri, S., Morita, K.-I., Kawaguchi, N., \& Hayakawa, M. 1997, \pasj, 49,
  123

\bibitem[{Lu {et~al.}(2014)Lu, Broderick, Baron, Monnier, Fish, Doeleman, \&
  Pankratius}]{Lu_etal.2014.apj}
Lu, R.-S., Broderick, A.~E., Baron, F., {et~al.} 2014, \apj, 788, 120

\bibitem[{Lu {et~al.}(2012)Lu, Fish, Weintroub, Doeleman, Bower, Freund,
  Friberg, Ho, Honma, Inoue, {et~al.}}]{Lu_etal.2012.apjl}
Lu, R.-S., Fish, V.~L., Weintroub, J., {et~al.} 2012, \apjl, 757, L14

\bibitem[{Lu {et~al.}(2013)Lu, Fish, Akiyama, Doeleman, Algaba, Bower,
  Brinkerink, Chamberlin, Crew, Cappallo, {et~al.}}]{Lu_etal.2013.apj}
Lu, R.-S., Fish, V.~L., Akiyama, K., {et~al.} 2013, \apj, 772, 13

\bibitem[{Mo{\'s}cibrodzka {et~al.}(2016)Mo{\'s}cibrodzka, Falcke, \&
  Shiokawa}]{Moscibrodzka_Falcke_Shiokawa.2016.aap}
Mo{\'s}cibrodzka, M., Falcke, H., \& Shiokawa, H. 2016, A\&A, 586, A38

\bibitem[{Rogers {et~al.}(1974)Rogers, Hinteregger, Whitney, Counselman,
  Shapiro, Wittels, Klemperer, Warnock, Clark, \&
  Hutton}]{Rogers_etal.1974.apj}
Rogers, A., Hinteregger, H., Whitney, A., {et~al.} 1974, \apj, 193, 293

\bibitem[{Rogers {et~al.}(1995)Rogers, Doeleman, \&
  Moran}]{Rogers_Doeleman_Moran.1995.aj}
Rogers, A.~E., Doeleman, S.~S., \& Moran, J.~M. 1995, \aj, 109, 1391

\bibitem[{Shepherd(1997)}]{Shepherd.1997}
Shepherd, M. 1997, G. Hunt \& HE Payne (San Francisco, CA: ASP), 77

\bibitem[{Tibshirani(1996)}]{Tibshirani.1996.jrssb}
Tibshirani, R. 1996, J. R. Statist. Soc. B, 58, 267

\end{thebibliography}

\end{document}